\documentclass{jaa}

\usepackage{natbib}
\usepackage{graphicx}

\begin{document}\sloppy

\title{TIRCAM2 Fast Sub-array Readout Mode for Lunar Occultation studies}


\author{Milind B. Naik\textsuperscript{1,*}, Devendra K. Ojha\textsuperscript{1},  Saurabh Sharma\textsuperscript{2},  Shailesh B. Bhagat\textsuperscript{1}, Savio L. D'Costa\textsuperscript{1}, Arpan Ghosh\textsuperscript{2} and Koshvendra Singh\textsuperscript{1}}

\affilOne{\textsuperscript{1}Department of Astronomy and Astrophysics, Tata Institute of Fundamental Research (TIFR), Mumbai 400005, India.\\}
\affilTwo{\textsuperscript{2}Aryabhatta Research Institute of Observational Sciences (ARIES), Manora Peak, Nainital 263001, India.}


\twocolumn[{

\maketitle

\corres{mbnaik@tifr.res.in}

\msinfo{1 July 2021}{1 July 2021}

\begin{abstract}
The TIFR Near Infrared Imaging Camera-II (TIRCAM2) is being used at the Devasthal Optical Telescope (DOT) operated by Aryabhatta Research Institute of Observational Sciences (ARIES), Nainital, Uttarakhand, India. In addition to the normal full frame observations, there has been a requirement for high speed sub-array observations for applications such as lunar occultation and star speckle observations. Fast sub-array modes have been implemented in  TIRCAM2 with suitable changes in the camera software at the computer and controller DSP code level. Successful observations have been carried out with the fast sub-array mode of observation. 
\end{abstract}


}]


\doinum{12.3456/s78910-011-012-3}
\artcitid{\#\#\#\#}
\volnum{000}
\year{0000}
\pgrange{1--}
\pgrange{1-5}
\setcounter{page}{1}
\setcounter{page}{1}
\lp{1}

\section{Introduction}

The TIFR Near Infrared Camera version 2 (TIRCAM2), is a near-infrared (NIR) astronomy camera \citep{2012BASI...40..531N,2018JAI.....750003B} which uses the Alladin-III Raytheon InSb detector of 512 x 512 pixel format. Though the detector is sensitive upto 5 $\mu$m, the optics of the camera limits the wavelength band from 1 to 3.7 $\mu$m. TIRCAM2 has been used at the focal plane of three telescopes in India: 1.2 m Mount Abu InfraRed Observatory (MIRO), Mount Abu (Physical Research Laboratory), 2 m IUCAA Girawali Observatory (IGO), Pune (Inter-University Center for Astronomy and Astrophysics) and 3.6 m Devasthal Optical Telescope (DOT), Devasthal (Aryabhatta Research Institute of Observational Sciences; ARIES). At present, it is mounted on the side port of the DOT. Since it is permanently mounted on the side port of DOT, it allows for near simultaneous observations in the NIR band alongwith the observations using the instrument mounted on the main port of the telescope.  TIRCAM2 is used for NIR observations with filters like J, H, K, Bracket $\gamma$, Polycyclic Aromatic Hydrocarbon (PAH) and narrow L-band. The fast sub-array imaging capability of TIRCAM2 is used primarily for lunar occultation observations which is useful for estimation of angular diameter of nearby stars and study of angular separation of binary stars as well as for star speckle observations.

\section{TIRCAM2 at DOT}

Initially, TIRCAM2 used to be mounted on the main port of the 3.6 m DOT as shown in Figure 1. Due to the increase in main port instruments and the opportunity to make near simultaneous observations in the NIR bands, the mounting arrangements for TIRCAM2 were modified to fit on the side port of DOT where it is permanently mounted during the observation season as shown in Figure 2. The side port arrangement shown in Figure 3 consists of the TIRCAM2 camera and the instrument rack which holds the power supplies, the control PC and all the electronics for the camera. All TIRCAM2 operations are carried out remotely from a laptop in the control room. The TIRCAM2 is mounted such that the focal plane of the telescope is just inside the window of the TIRCAM2 cryostat. The focal plane is reimaged using lenses on to the InSb detector array through a filter wheel. The filters can be changed by the operator using the filter controller software on the control PC in the rack which can be accessed remotely from the control room laptop. The detector array is controlled by an Astronomical Research Camera (ARC) controller (also known as Leach controller) shown in Figure 4. Data and commands between the control PC and the ARC controller are sent over an Optical Fibre Cable (OFC) pair. A block diagram of the ARC controller is shown in Figure 5. The TIRCAM2 optical elements are maintained at around a temperature of 17 K and the InSb detector at around a temperature of 35 K with the help of a closed-cycle Helium gas CTI cryo-cooler. The temperature of the Focal Plane Array (FPA) detector is controlled by a Lakeshore temperature controller. The two Helium gas pipes from the compressor to the cold head mounted on the TIRCAM2 cryostat are routed through the telescope pier.

\begin{figure}[!t]
\includegraphics[width=.8\columnwidth]{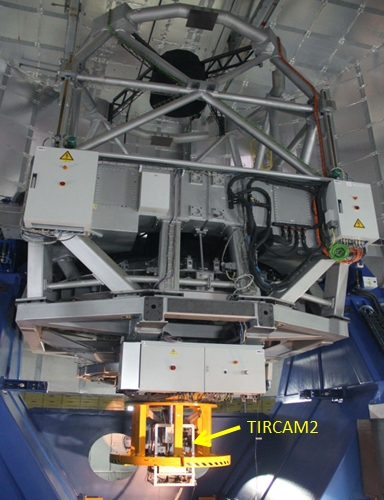}
\caption{TIRCAM2 mounted at DOT main port.}\label{figOne}
\end{figure}

\begin{figure}[!t]
\includegraphics[width=.8\columnwidth]{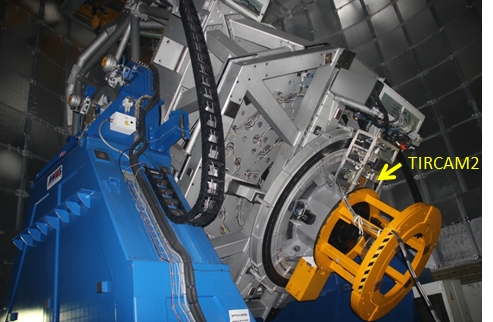}
\caption{TIRCAM2 observational setup at DOT side port.}\label{figTwo}
\end{figure}

\begin{figure}[!t]
\includegraphics[width=.8\columnwidth]{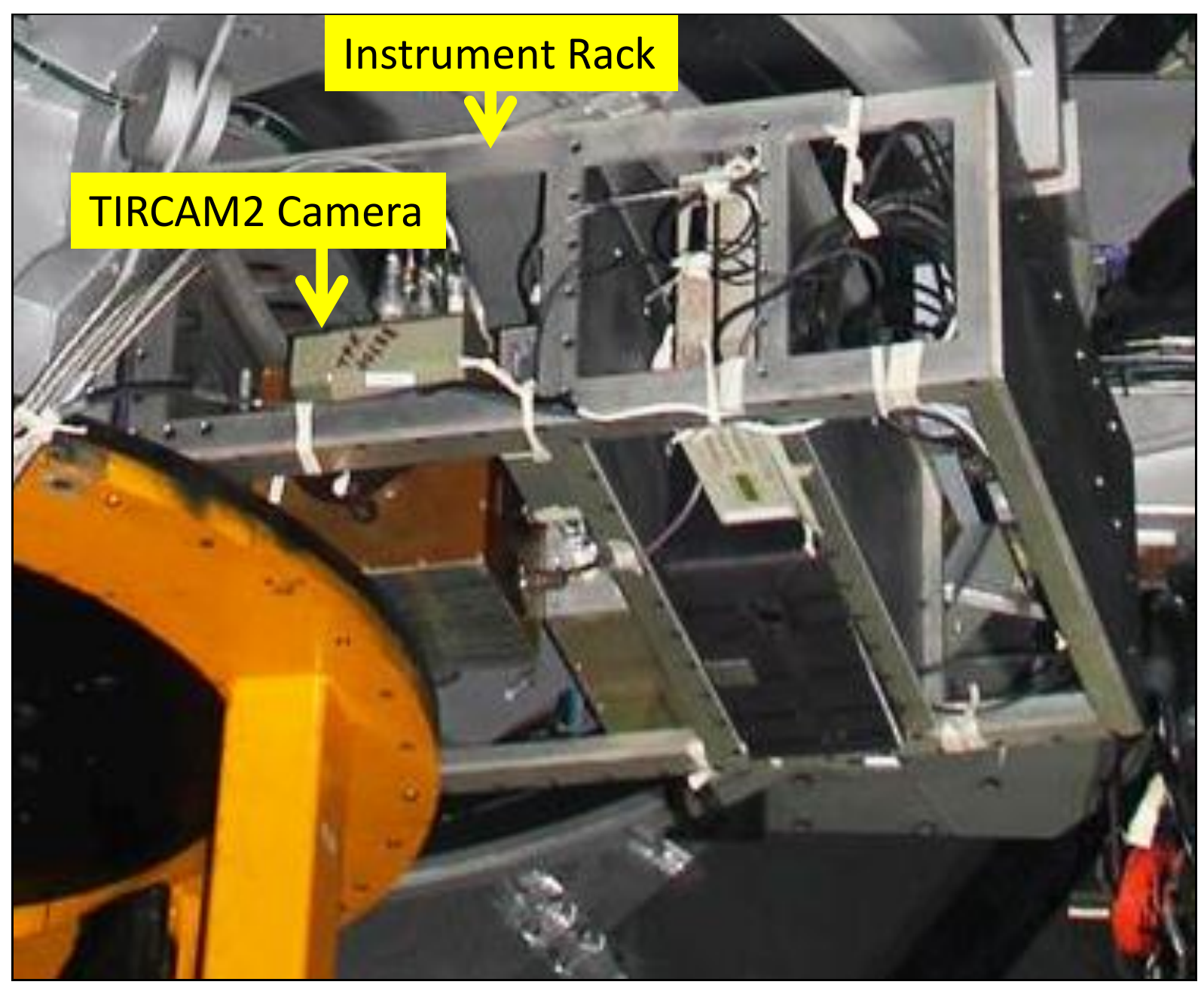}
\caption{TIRCAM2 instrument setup at DOT side port.}\label{figThree}
\end{figure}

\section{TIRCAM2 electronics}

The TIRCAM2's FPA is controlled using an ARC controller shown in Figure 4. The controller consists of a Timing board with an onboard DSP chip, a Clock Board and a Video board mounted in a six slot controller chassis. The timing board communicates with the control PC through a PCI/PCIe card mounted in the control PC through a pair of optical fibre cables.

\begin{figure}[!h]
\includegraphics[width=.65\columnwidth]{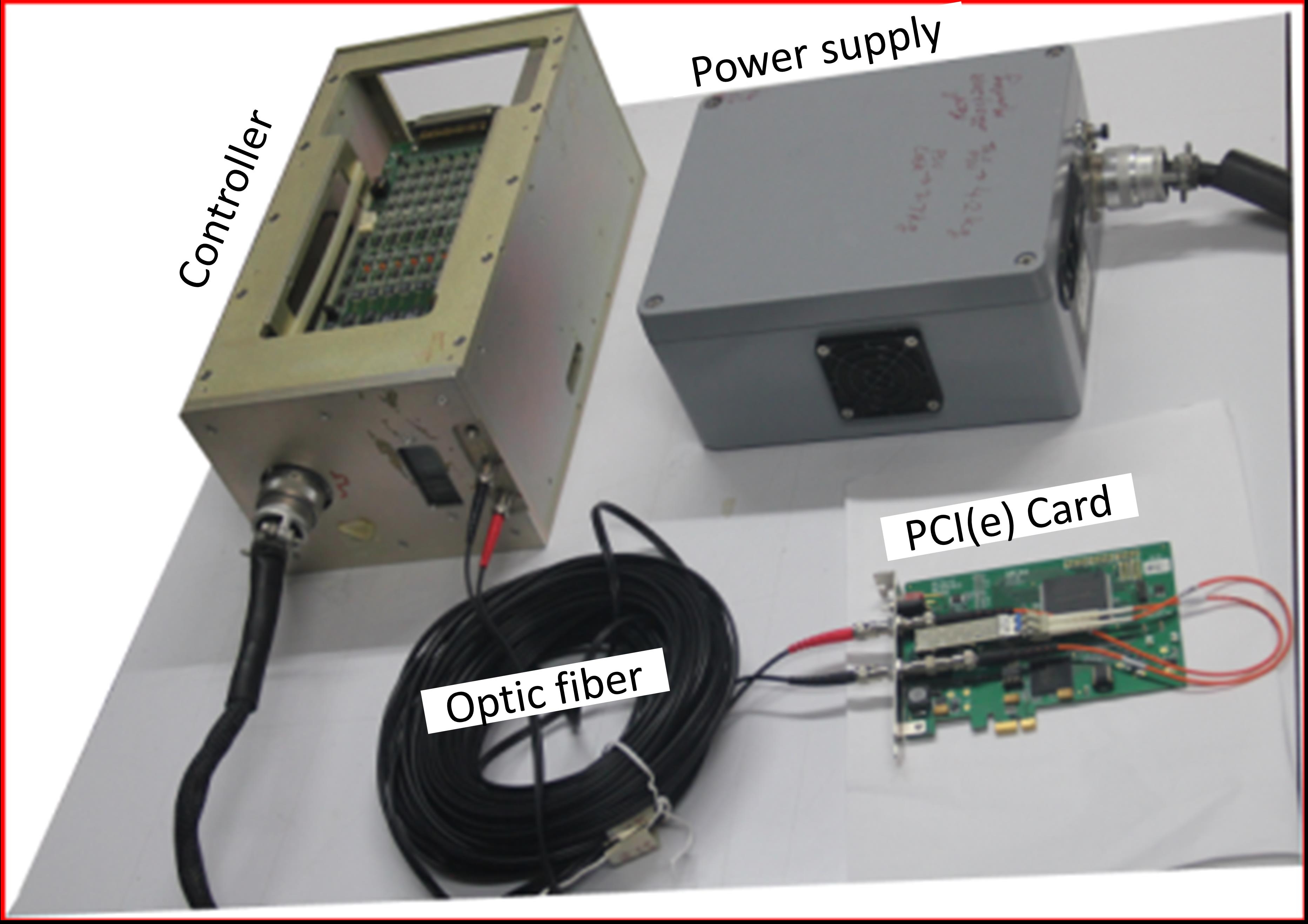}
\caption{ARC controller (Leach controller).}\label{figFour}
\end{figure}

The different blocks of the ARC Controller are shown in Figure 5. The block on the left shows the control computer with the PCI/PCIe card and the software interface. The Voodoo Graphical User Interface (GUI), written in Java, communicates with the API library to carry out system and PCI card operations like sending commands to the Controller cards and receiving, processing and storing image data from the TIRCAM2.

\begin{figure}[!h]
\includegraphics[width=1\columnwidth]{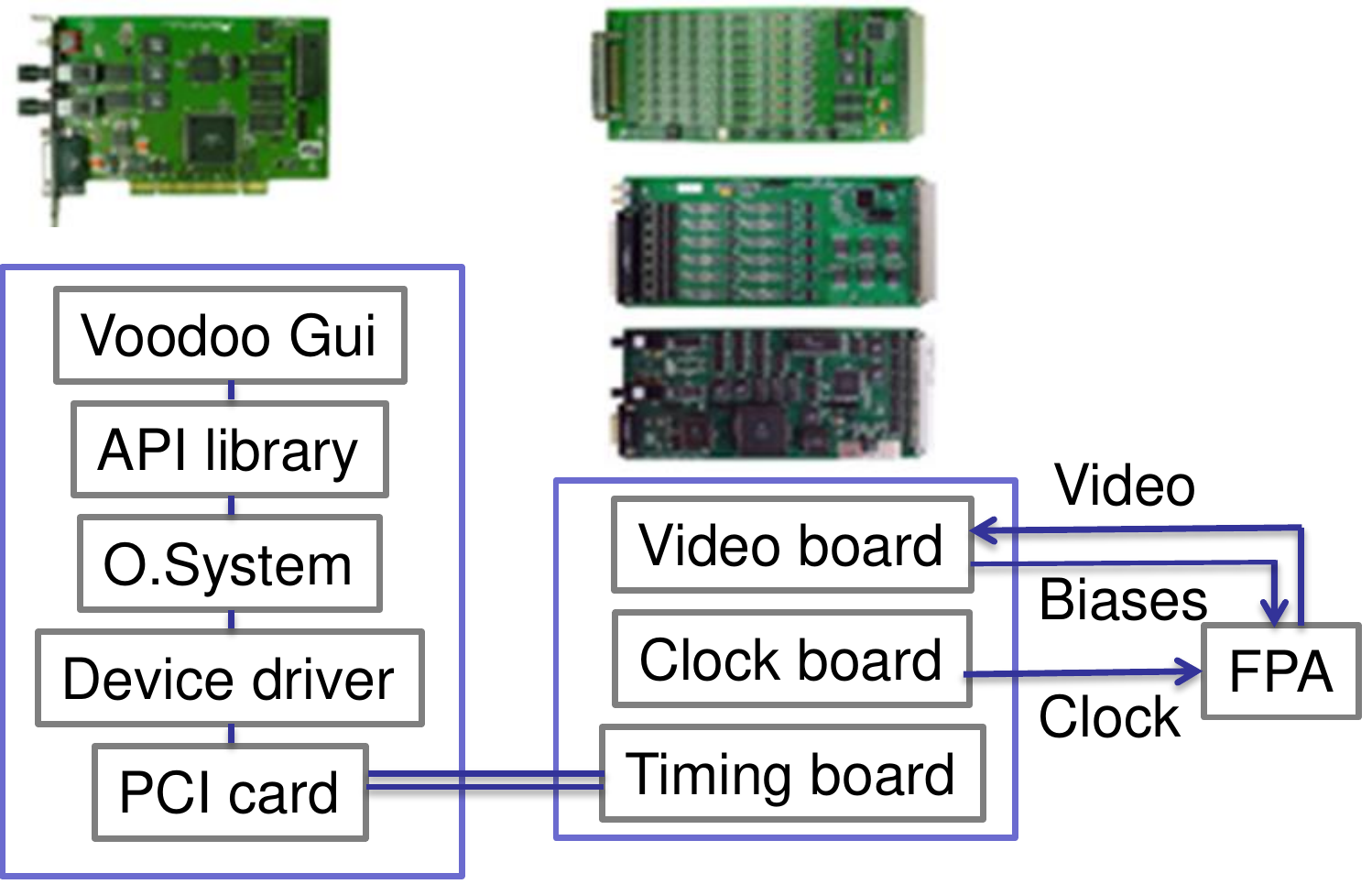}
\caption{ARC controller block diagram }\label{figFive}
\end{figure}

The block on the right shows the controller hardware which consists of three boards: the Timing board, the Clock board and the Video board which are physically mounted on a six slot chassis with a backplane. The Timing board has a DSP chip which controls the overall operations of the TIRCAM2 FPA and has its code written in Assembly language. The Clock board generates the clocking pattern and bias voltages required to operate the FPA. The Video board has 8 video channels which process and digitise the image output signals from the FPA and it can also provide 7 bias voltages.

\section{Data capture}

The main data capture screen of the Voodoo GUI is shown in Figure 6. The user inputs the FITS file path and name, the NIR filter selected for observation, the exposure time and the number of frames to capture. Data capture commences when the Expose button is clicked and the required number of frames are read. The full frame of the FPA has 512 x 512 pixels and the full frame capture time is about 256 ms.

\begin{figure}[!h]
\includegraphics[width=.8\columnwidth]{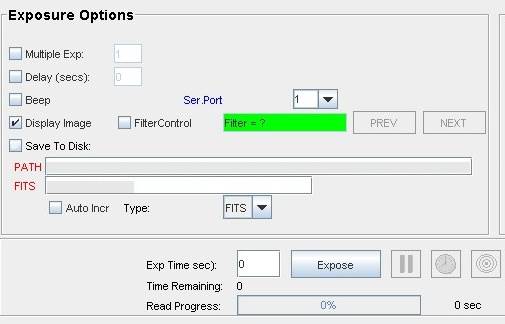}
\caption{Main data capture screen}\label{figFour}
\end{figure}

In the case of fast sub-array readout, another dialog box is opened and the user inputs the Box Width, the Box Height, the Box Centre Column and the Box Centre Row as shown in Figure 7. This information is sent to the controller over OFC. In this mode of readout, the continuous capture mode is used.  

\begin{figure}[!h]
\includegraphics[width=.8\columnwidth]{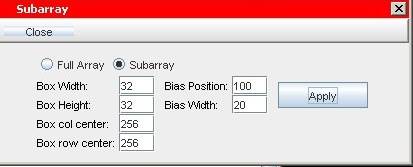}
\caption{Setting for sub-array box on FPA.}\label{figFour}
\end{figure}

The Voodoo GUI Java code and the Controller Assembly Code were suitably modified to use the Continuous Capture Mode of the Controller. The GUI input screen for the Continuous Mode is shown in Figure 8. At present, the modified code supports capture of upto 4000 frames. The Voodoo software calculates the number of frames that can be grabbed which is limited by the size of the system memory buffer and sends the number `$MAX\_FRAME\_BUFFER$'  to the controller. It also sends the number of frames requested (4000 in this case) as `$REQ\_FRAMES$' to the controller.

\begin{figure}[!h]
\includegraphics[width=.8\columnwidth]{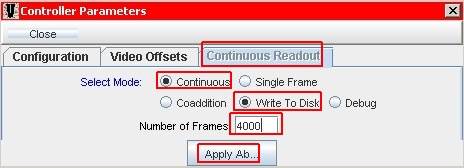}
\caption{Set up of parameters for continuous capture mode of controller.}\label{figFour}
\end{figure}

The sub-array frames are captured after the Expose button is clicked and the progress is displayed on the progress bar as shown in Figure 9.

\begin{figure}[!h]
\includegraphics[width=.8\columnwidth]{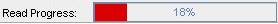}
\caption{ Read progress bar indicator.}\label{figFour}
\end{figure}

A single FITS file with 4000 sub-array frames is generated after a successful sub-array capture. A time stamp is added in the FITS header as `start time' and `end time'. The Java code for FITS header routine which earlier had a resolution of 1 s was modified to get a 1 ms resolution time stamp. A standalone program was written to convert the FITS file of 4000 frames into a data cube FITS file with 4000 frames as shown in Figure 10. The size of data cube is about 8 MB.

\begin{figure}[!h]
\includegraphics[width=.8\columnwidth]{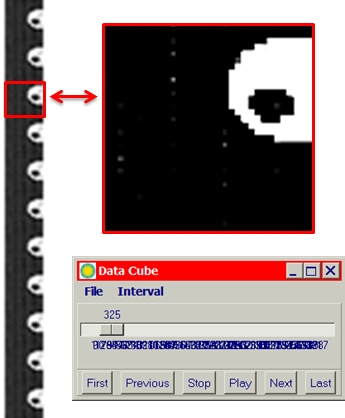}
\caption{ Screen shot of program to split 4000 frame FITS to single datacube FITS.}\label{figFour}
\end{figure}

\begin{figure}[!h]
    \centering
    \includegraphics[height=.40\textheight]{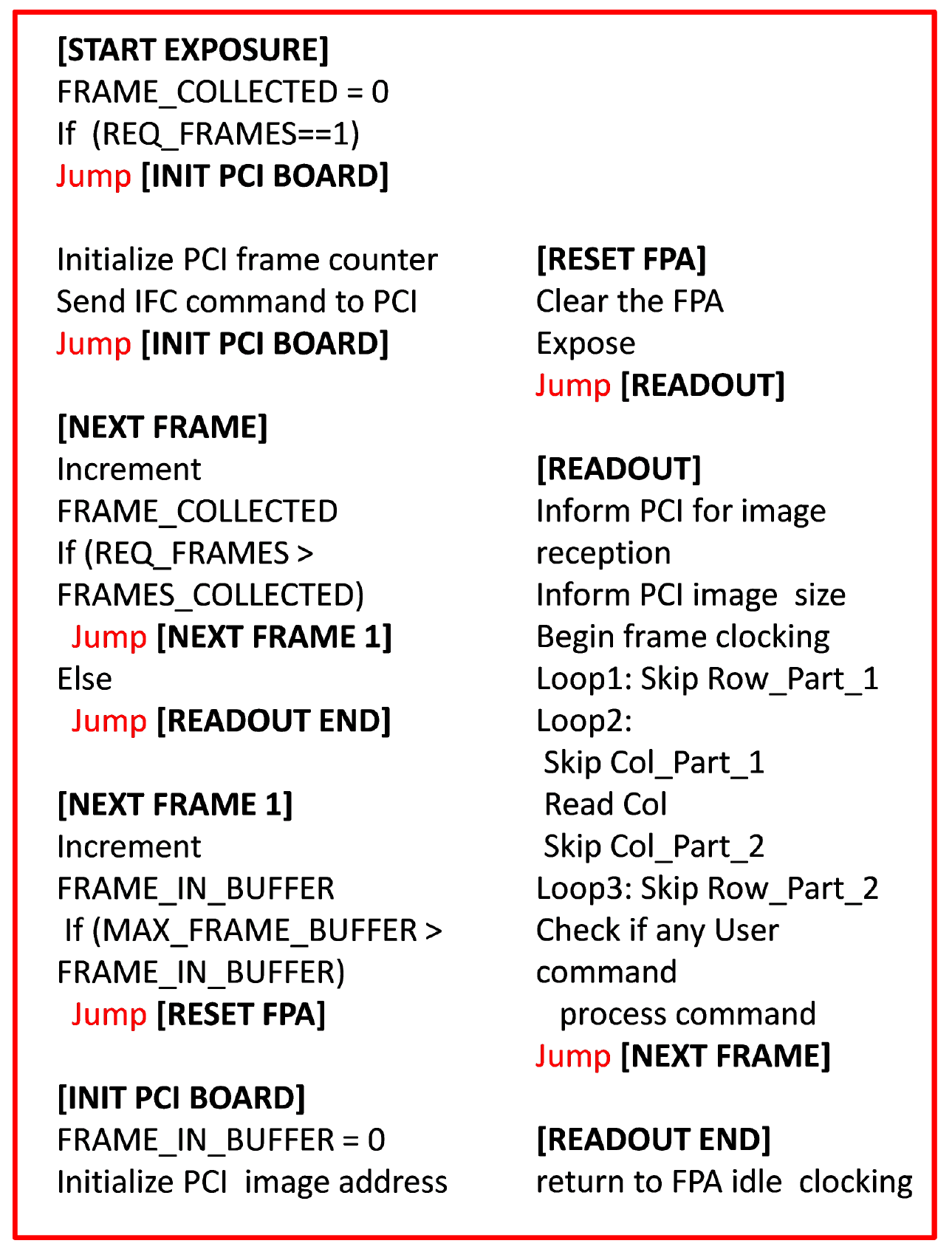}
    \caption{DSP pseudo code for sub-array.}
    \label{fig:my_label}
\end{figure}

\section{Software}

The Controller DSP assembly language code was appropriately changed for sub-array mode. Figure 11 shows the DSP pseudo code of the relevant section. The computer memory buffer allows storage of only certain number of frames which is calculated and provided by the Voodoo Java code in the form of `$MAX\_FRAME\_BUFFER$'. The number of frames requested is provided by the variable `$REQ\_FRAMES$'. After the Expose button is clicked, the sequence of operations for capturing the required number of frames assuming a 32 x 32 box is required is as follows and is shown in Figure 12:
\begin{enumerate}
    \item FPA is reset and all charges collected are cleared  (see Figure 13).
    \item Skip Rows 1 till the 32x32 box is reached.
    \item Capture/Digitize box and send data to PC after skipping the unwanted columns.
    \item Skip Rows 2 beyond the box.
\end{enumerate}

After the required number of frames are captured, the frame data stored in the PC memory buffer are written to a FITS file.
The time between two sub-array frames is about 9.9 ms (see Figure 13) and capture box light integration timing is about 4.5 ms.

\begin{figure}[H]
\includegraphics[height=.33\textheight]{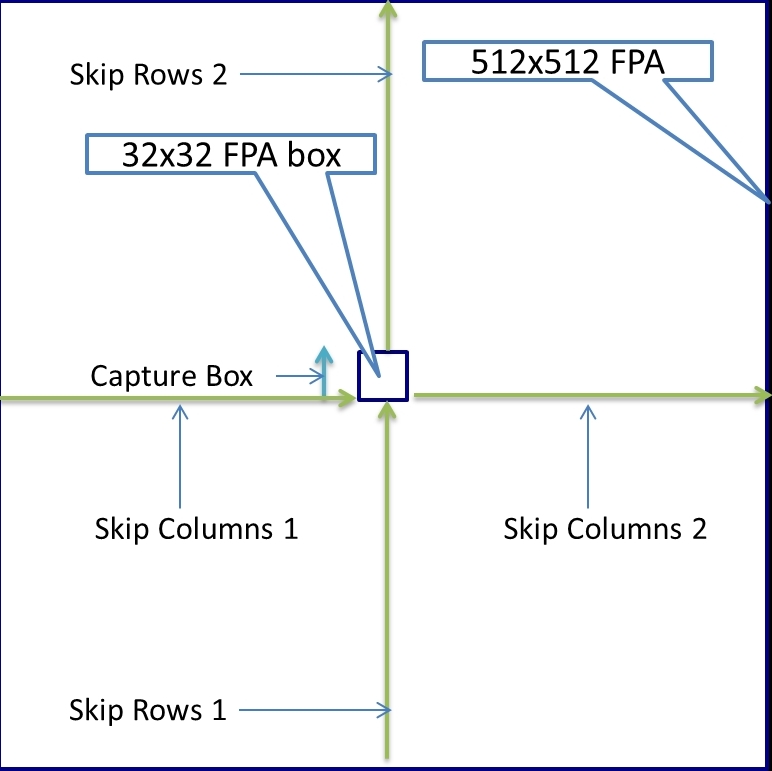}
\caption{FPA capture box and row/column skip mechanism. }\label{figEleven}
\end{figure}

\begin{figure}[!h]
\includegraphics[height=0.23\textwidth ]{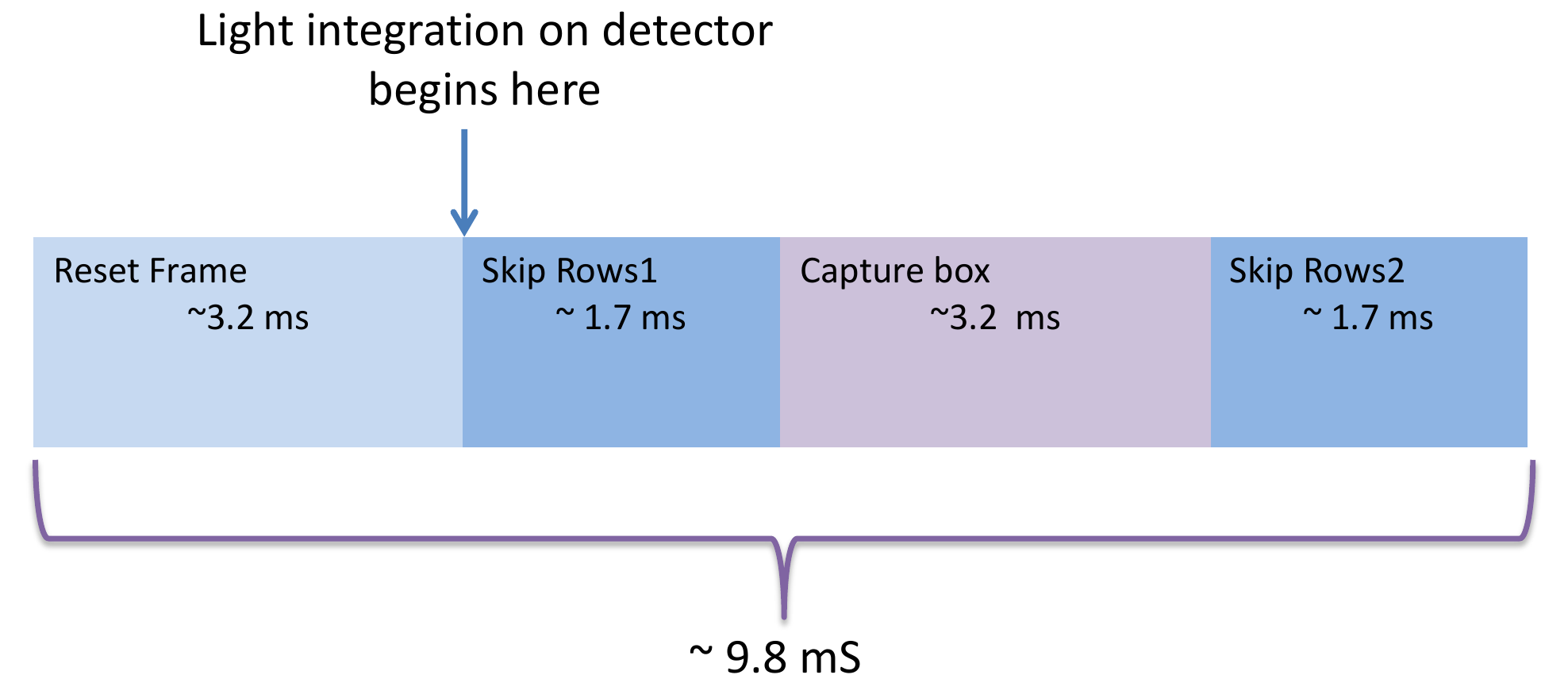}  
\caption{Box size versus sampling time is shown. It takes $\sim$ 1.7 ms to skip initial part of FPA (Skip Rows 1) till capture box. Then Capture box [32x32 pixels] is digitized in $\sim$ 3.2 ms and then rest of rows are skipped (Skip Rows2).}\label{figFive}
\end{figure}

\section{Observation Results}

The observation of IRC +20101 M0 giant where star speckle could be seen using the sub-array readout mode is shown in Figure 14. The light curves shown in Figure 15 were generated using pre-existing software described in \citet{2020MNRAS.498.2263R}

\begin{figure}[!h]
\includegraphics[height=.25\textheight]{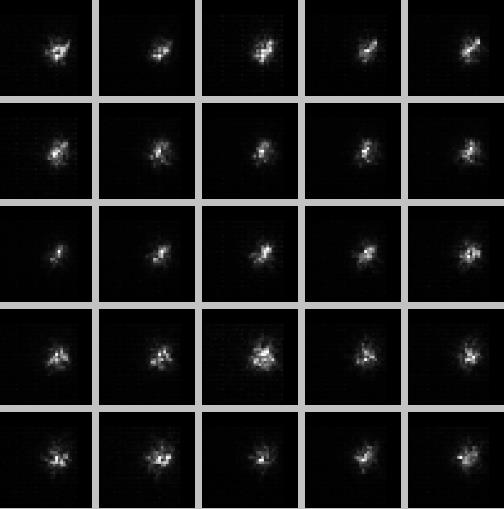}
\caption{Sample of star speckle obtained with TIRCAM2 sub array (32x32 pixel)  mode at DOT.}
\end{figure}

\begin{figure}[!h]
\includegraphics[height=.15\textheight]{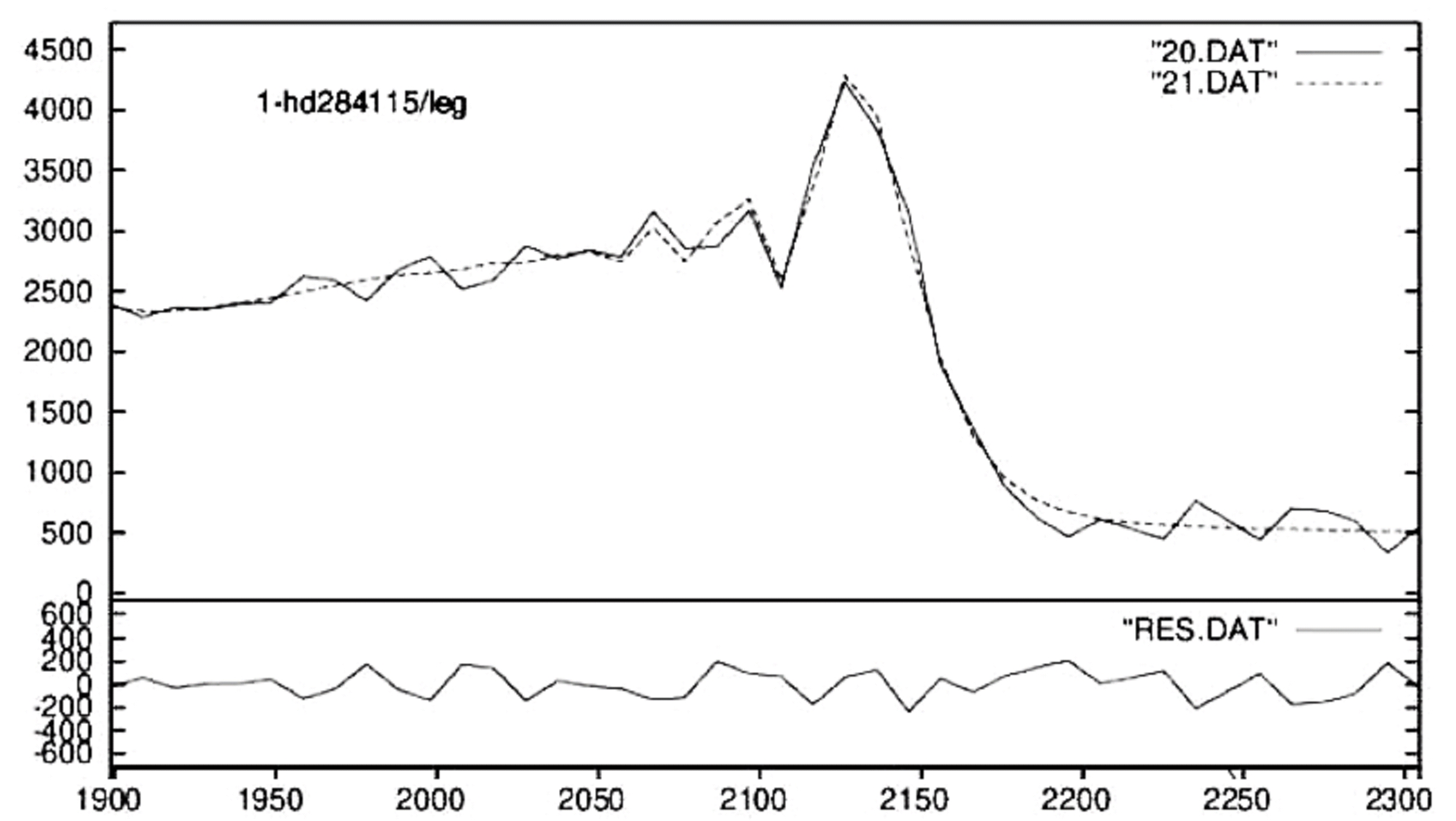}
\caption{A sample of lunar occultation light curve obtained with TIRCAM2 sub-array (32x32 pixel)  mode at DOT. 20.DAT and 21.DAT are the data points and fitted line respectively.}
\end{figure}

\vspace{-2em}

\section*{Acknowledgements}
The authors would like to thank all members of ARIES/DOT for their support in operating TIRCAM2 during the observations. We would also like to thank Dr. Andrea Richichi for guidance and inputs which made these observations a success. We thank the personnel at the TIFR Central Workshop for fabricating the telescope interfaces required for mounting TIRCAM2 on DOT. The continuous help and support of all the members of the TIFR Infrared Astronomy Group and the Department of Astronomy and Astrophysics is highly appreciated. M.B.N., D.K.O., S.B.B., S.L.D. and K.S. acknowledge the support of the Department of Atomic Energy, Government of India, under Project Identification No. RTI 4002.\\

\vspace{-1em}

\bibliography{reference}

\vspace{-1.5em}

\end{document}